\pdfoutput=1
\documentclass{article}

\usepackage{arxiv}
\usepackage[T1]{fontenc}
\usepackage{newtxtext,newtxmath}
\usepackage{microtype}
\usepackage{graphicx}
\usepackage{booktabs}
\usepackage{xcolor}
\usepackage[round,sort&compress]{natbib}
\usepackage[font=small,labelfont=bf,labelsep=period]{caption}
\usepackage{enumitem}
\usepackage[hidelinks]{hyperref}
\usepackage{doi}
\usepackage{orcidlink}

\definecolor{marsblue}{HTML}{2F6690}
\hypersetup{colorlinks,linkcolor=marsblue,citecolor=marsblue,urlcolor=marsblue}

\newcommand{\figslot}[2]{\includegraphics[width=\textwidth]{figures/#1}}

\renewcommand{\headeright}{Preprint}
\renewcommand{\undertitle}{Preprint}
\renewcommand{\shorttitle}{MarsNet}

\title{\texorpdfstring{MarsNet\kern0.05em:}{MarsNet:} learning the Martian atmosphere from observations alone}

\author{%
  \orcidlink{0000-0003-3305-5924}\,Sebastian Mengwall\thanks{Correspondence: \href{mailto:sebastian.mengwall@yale.edu}{sebastian.mengwall@yale.edu}}\\
  Yale University\\
  New Haven, CT, USA\\
  \And
  \orcidlink{0000-0003-3943-5194}\,J.~Michael Battalio\\
  Department of Earth and Planetary Sciences\\
  Yale University\\
  New Haven, CT, USA\\
}

\date{}

\begin{document}
\maketitle

\begin{abstract}
We present MarsNet, a machine-learned model of the Martian atmosphere trained only on observations. Existing reanalyses assimilate spacecraft retrievals into general circulation models (GCMs), so where observations are sparse their fields largely reflect the model. We therefore use extant reanalyses only for evaluation. MarsNet alternates a two-hour forecast with an update that assimilates the observations of each window, and both steps are learned jointly. Because the Martian record is too sparse for an unrestricted network, learning is confined within a fixed physical structure: the state is in physical units, the forecast uses physical tendencies of fixed form and learned strength, and the update is a bounded Kalman update. MarsNet is trained on Mars Years (MY) 24--31 and evaluated on independent observations. On three seasons of MGS radio-occultation temperatures assimilated by no dataset, MarsNet matches or improves on the three reanalyses in the lowest three scale heights, with a near-surface root-mean-square error (RMSE) of 4.2~K against 4.4--5.4~K, but is less accurate aloft. The model's surface pressures are the most accurate of the four products (RMSE 10~Pa against Curiosity in MY32, compared with 20--46~Pa). On withheld Mars Climate Sounder profiles of MY32, its temperatures beat an observed climatology and persistence, including in ten-sol forecasts without new observations, and its stated uncertainty is close to calibrated. Much of the temperature skill is present before training and appears to come from the structure and tables fitted to the observations, while training mainly improves the surface pressures, at some cost to dust. The model cycles stably through nine Mars years, and the influence of each observation can be traced. Its speed makes experiments on the observing record practical, and it offers an independent reference for GCMs and reanalyses.
\end{abstract}


\section{Introduction}

Three reanalyses represent the Martian atmosphere over the past twenty Earth years using assimilated observations. The Mars Analysis Correction Data Assimilation \citep[MACDA,][]{montabone2014macda, valeanu2026macda} and Open access to Mars Assimilated Remote Soundings \citep[OpenMARS,][]{holmes2020openmars} assimilate Thermal Emission Spectrometer (TES) onboard Mars Global Surveyor (MGS) and Mars Climate Sounder (MCS) onboard Mars Reconnaissance Orbiter (MRO) retrievals into the UK version of the LMD Mars GCM \citep{forget1999lmd} with the Analysis Correction scheme \citep{lewis2007ac}. Version 2.0 of MACDA also assimilates retrievals from the Thermal Emission Imaging System (THEMIS) onboard Mars Odyssey \citep{smith2009themis}. The Ensemble Mars Atmosphere Reanalysis System \citep[EMARS,][]{greybush2019emars} assimilates temperatures from TES and MCS into the GFDL Mars GCM with an ensemble Kalman filter \citep{hunt2007letkf}. These products are widely used for Mars weather analysis, climate science, and mission planning.

\begin{figure}[t]
\centering
\figslot{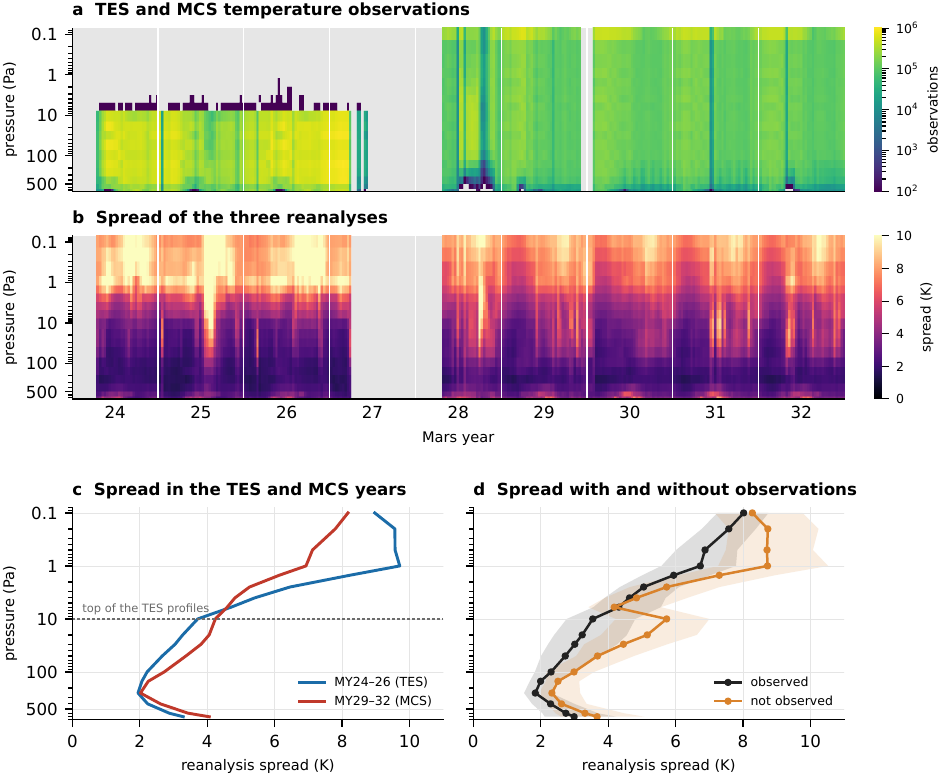}{temperature observations and the disagreement of the three reanalyses, MY24--32.}
\caption{The density of temperature observations is compared with the spread among the reanalyses, to show where the observations constrain them. (a) TES and MCS temperature observations per 10$^\circ$ of $L_s$ and per scale height of pressure, summed over latitude. (b) The standard deviation of the temperatures of MACDA, OpenMARS, and EMARS, averaged over 10$^\circ$ of $L_s$ and over latitude with cosine weights; gray where the three do not all exist. (c) The mean spread against pressure in the TES years MY24--26 and the MCS years MY29--32; the dashed line marks the top of the TES profiles, while MCS reaches the top of the panel. (d) The median and interquartile range of the spread, at each level, over bins of 10$^\circ$ of $L_s$ and 5$^\circ$ of latitude with at least 100 observations per scale height (observed) and with none.}
\label{fig:constraint}
\end{figure}

Between observations, and for quantities that are rarely or never observed, the fields of a reanalysis come from the dynamical core, the dust and cloud parameterizations, and the boundary layer of its GCM, so a reanalysis can be no more independent than its model wherever observations are sparse. In such regions, the three reanalyses often disagree \citep[e.g.,][]{Battalio2022}. Figure~\ref{fig:constraint} shows this for temperature over MY24--32. The three products agree to about 2~K near 250~Pa, which both TES and MCS observe, but differ by up to 10~K at pressures below 10~Pa in MY24--26, when the TES nadir profiles reached only to about 10~Pa. MCS retrievals extend to 0.1~Pa, so the disagreement at 1~Pa falls from 9.7 to 6.9~K, while near the surface, which the MCS limb views reach less often, it rises from 3.3 to 4.1~K. Even where observations are dense, the products still differ by several kelvin at pressures less than 10~Pa. Nor is the disagreement limited to temperature. Against Curiosity's surface pressures in MY32, which no dataset assimilates, the mean bias ranges from $-37$ to $+28$~Pa (Section~\ref{sec:results}). Reanalyses are also usually checked against the same observations they assimilate, and such a check cannot establish which product is more accurate. As a result, no existing reanalysis is an obvious ``true'' state on which to train a machine-learned model of Mars.

On Earth, machine-learned models (graph neural networks) now forecast the weather about as well as physics-based models, depending on the situation and research question \citep{lam2023graphcast, price2025gencast}. Most of these models are trained on reanalysis fields, so they inherit the reanalysis. Recent models also take in observations, either alongside a reanalysis, as in Aardvark and WeatherNext 3 \citep{allen2025aardvark, rasp2026weathernext3}, or alone, as in GraphDOP \citep{alexe2024graphdop}, which has also produced a multi-decade reanalysis of Earth's atmosphere \citep{lean2026obsreanalysis}. However, new observation types introduced into a model trained on a reanalysis can conflict with the training data \citep{yuval2026neuralgcm}.

On Mars, machine learning has been used to identify water-ice clouds in daily global maps \citep{mengwall2023clouds}, and the first machine-learned models of the atmosphere are trained on the gridded fields of a reanalysis or of the Mars Climate Database and judged against those fields \citep{roy2026marsfm, singh2026aircastmars, carroll2026marscast}.

Mars has far fewer observations than Earth, so a large model left free to learn from them has far more degrees of freedom than the observations can constrain. However, a network trained only on observations from a single infrared instrument, much like the situation on Mars, can still infer large-scale features such as jet streams and surface pressure patterns \citep{lean2025graphdop}. We therefore build in the structure that classical data assimilation already provides and allow learning to adjust it only where the observations constrain it. A fitted climatology holds the recurring seasonal and daily cycles, linear models carry the large-scale memory, and learning sets the strength of named physical processes and how strongly each observation corrects the state. The forecast and the update are trained together to predict each observation before it is assimilated. In return, the effect of each measurement on the estimate can be followed and tested. The estimate does not depend on any GCM run, although it inherits any model assumptions that the retrievals themselves contain (Section~\ref{sec:data}).

A model trained on observations must also be tested on observations it has not used, which requires care on Mars, because every assimilated temperature passes through a correction of instrument biases and that correction has to be referred to some instrument. We refer it to the MGS radio occultations and hold out whole seasons of them, together with all Mars Express (MEX) occultations, from the assimilation, from training, and from everything fitted to observations before training. These occultations are the only temperatures that neither MarsNet nor any reanalysis has used, so they are the benchmark for comparing the products. Against them, MarsNet matches or improves on the reanalyses near the surface but is less accurate higher up, and its surface pressures are the most accurate of the four products. The validation year, MY32, tests MarsNet itself, as a function of the time since it last assimilated an observation.

Section~\ref{sec:data} describes the observations and how they are divided between training, assimilation, and testing, Section~\ref{sec:method} details the model, Section~\ref{sec:results} quantifies its skill on held-out observations and what each observation does to it, and Section~\ref{sec:discussion} highlights what MarsNet offers, its limits, and its uses. The appendices give detail on the observations and the model, the instrument temperature correction, training, and stability tests.

\section{Observations and evaluation design}
\label{sec:data}

MarsNet assimilates geolocated observations, each with a time, a position, and a stated error. The assimilated inputs are MCS limb profiles of temperature, dust extinction, and water-ice extinction \citep{mccleese2007mcs,kleinbohl2009mcs}, including the heights at which the retrieval reports only an upper limit of the aerosol extinction; TES nadir temperature profiles and dust and water-ice columns \citep{christensen2001tes, conrath2000tes, smith2004tes}; and column opacity from the Mars Exploration Rovers Spirit, in Gusev Crater (14.5$^\circ$S, 175.5$^\circ$E), and Opportunity, in Meridiani Planum (1.9$^\circ$S, 354.5$^\circ$E), and from the Mars Science Laboratory rover Curiosity, in Gale Crater (5.3$^\circ$S, 137.8$^\circ$E) \citep{lemmon2015opacity, msl_opacity_lemmon2024}. We use version 6.2 of the MCS profiles in the Planetary Data System \citep{mccleese2025mcsddr}, a later retrieval than versions 5 and 5.2 assimilated by EMARS, MACDA, and OpenMARS.

Some observations are not assimilated but are scored at their positions as training targets. These include the MCS dust and water-ice column opacity, the MCS and TES surface temperatures, and upper-atmosphere temperatures from stellar occultations by the Spectroscopy for the Investigation of the Characteristics of the Atmosphere of Mars (SPICAM) on MEX \citep{forget2009spicam}. Water-vapor columns from the Compact Reconnaissance Imaging Spectrometer for Mars (CRISM) \citep{toigo2013crism,smith2018crism} are also targets, and those of the training years set the water vapor of the model's climatology (Section~\ref{sec:state}). Unassimilated surface observations include the surface and air temperatures of Curiosity's Rover Environmental Monitoring Station (REMS) \citep{gomezelvira2012rems}. Surface pressures are not assimilated individually. Those of the Viking landers in MY12--15 \citep{tillman1989viking}, REMS in MY31 \citep{harri2014rems}, and the Phoenix lander in MY29, in Vastitas Borealis (68.2$^\circ$N, 234.3$^\circ$E) \citep{taylor2008phoenix, taylor2010phoenix}, fix the seasonal cycle and daily tide of the model's surface pressure, and the lander and occultation pressures outside the held sets correct its slowly varying bias field (Section~\ref{sec:state}). Surface pressures from REMS, Phoenix, and the MGS radio occultations \citep{hinson1999mgsrs} also serve as training targets.

MarsNet uses fixed maps of the surface: Mars Orbiter Laser Altimeter (MOLA) topography and roughness \citep{smith2001mola} and TES albedo and thermal inertia \citep{christensen2001tes, putzig2007ti}. Most observations are MCS and TES profiles (Figure~\ref{fig:coverage}).

We enforce two rules on assimilated data to keep the model observation-only. First, no reanalysis, GCM run, or gridded atmospheric product is ever used as an input, a training target, a normalization statistic, or a starting state. Second, every file comes from the primary instrument archive. Sometimes, the retrievals themselves carry some model information. For example, the MCS retrieval starts from a GCM climatology and fixes the surface-air temperature contrast at that climatology's value \citep{kleinbohl2009mcs}, so MCS temperatures in the lowest scale height are somewhat constrained by GCM dynamics.

\subsection{Evaluation design}
\label{sec:eval}

MarsNet's model trains on MY24--31 and is validated on MY32. MY33--34 are test years that nothing in this work reads, so no observation after MY32 is used for training or scoring in our initial version.

\begin{figure}[t]
\centering
\figslot{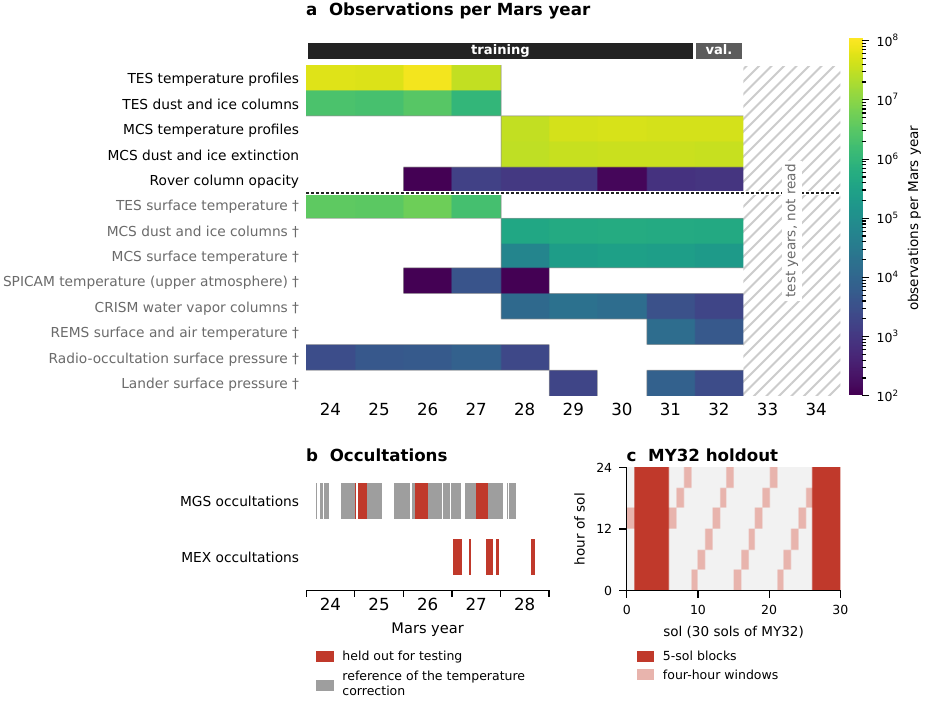}{observations per Mars year and kind, with the held sets shaded and a zoom on the MY32 blocks and slices.}
\caption{The observations are counted by year and kind, and the parts held out for testing are marked. (a) Observations per Mars year and kind. Each observation is one profile level, column, or surface value. (b) The occultation profiles in 5$^\circ$ bins of $L_s$: red bins are held out for testing (Table~\ref{tab:indep}), the MGS occultations of three seasons and every MEX occultation, and gray bins are the MGS reference of the temperature correction. (c) The MY32 holdout over 30 sols: one 5-sol block in every 25 sols and one 4-hour window per sol, rotating through the day. $\dagger$\,Observations below the dashed line are scored in training but never assimilated; the surface pressures also set the mean surface pressure (Section~\ref{sec:state}). The test years MY33--34 are not read.}
\label{fig:coverage}
\end{figure}

In the validation year, all observations of every instrument in one 5-sol block in every 25 sols and in one 4-hour window per sol, rotating through the day from sol to sol, are scored and never assimilated (Figure~\ref{fig:coverage}). Together these cover a third of the time of MY32. A profile in a 4-hour window has assimilated profiles of the same instrument within hours, so it tests the analysis. Within a 5-sol block, however, no observation of any instrument is assimilated anywhere on the planet, so a profile within the block can be up to five sols from the last assimilated observation, since the model uses only past observations. This tests how well the model carries the state through a gap with no data. We therefore report skill separately for the two and by the time since the last assimilated observation of the same instrument within 800~km, which can exceed five sols where the orbits had not passed nearby before a block.

Temperatures from different instruments disagree by a few kelvin, so we attempt to correct the TES and MCS temperatures for their biases before they are assimilated (Appendix~\ref{sec:tbias}). The correction needs a reference, and we use the MGS radio occultations, which measure temperature from the refraction of a radio signal and need no prior. A correction fitted on all occultations would leave none independent of it, so we hold out three whole seasons of MGS occultations, staggered by year ($L_s = 0$--$90^\circ$ of MY25, $90$--$180^\circ$ of MY26, and $180$--$270^\circ$ of MY27, for a total of 4,475 of 21,244 profiles), and all 301 MEX occultations of MY27--28 \citep{tellmann2013mars}. Each held season is longer than the seasonal resolution of the correction, so no fitted profile shares its season and year.

Besides the weights learned in training, MarsNet uses fitted tables, fixed quantities fitted to observations before training and not changed by it: the instrument temperature correction, the model's climatology, the Linear Inverse Models (LIMs) (Section~\ref{sec:state}), the datum of surface pressure, the scaling of the predicted spread of MCS temperatures (TES is not scaled, because its ratio of error to predicted spread varies too much between years for one fixed scaling to correct), and the relation between the pressure tide at Curiosity and the dust column (Appendix~\ref{app:model}). The held occultations are never assimilated, never training targets, and never used in any fitted table. Their surface pressures are held out with them. Outside the held seasons, the occultation surface pressures of one 5-sol block in every 25 sols are also held out for scoring.

To test dust, we withhold the column opacity of one rover site for each whole year (Opportunity in MY26, MY28, MY30, and MY32, Spirit in MY27 and MY29, and Curiosity in MY31), and 30\% of the MCS orbits during the MY28 global dust storm ($L_s = 262$--$310^\circ$) from training and from both the model's climatology and the reference climatology used for scoring (Sections~\ref{sec:state} and \ref{sec:results}).

Table~\ref{tab:indep} lists, for every scored set, each way it could have reached MarsNet or the reanalyses. All three reanalyses assimilate TES and MCS, so we compare the products only on the occultations and on Curiosity's surface pressures, which none of them assimilate.

\begin{table}[t]
\centering
\small
\caption{Each scored observation set is checked against every pathway by which it could reach MarsNet or the reanalyses, to show which sets are independent. The columns ask, in order, whether the observations of a set were assimilated by MarsNet, served as training targets, were used to fit one of the fitted tables (Section~\ref{sec:eval}), were used to pick the model, by stopping training, choosing among candidate runs, or deciding a change to the code, and were assimilated by a reanalysis. ``Partly'' means that not every reanalysis assimilates MCS dust. Curiosity's pressures outside the MY32 holdout correct the dust column through the pressure tide (Section~\ref{sec:update}), and the friction of the winds near the equator was chosen in comparison to Curiosity's observed winds.}
\label{tab:indep}
\begin{tabular}{lccccc}
\toprule
Set & Assimilated & \shortstack{Training\\target} & \shortstack{Fitted\\tables} & \shortstack{Used to pick\\the model} & Reanalyses \\
\midrule
\multicolumn{6}{l}{\textit{Independent tests}} \\
MGS occultation $T$, held seasons (MY25--27) & no & no & no & no & no \\
MEX occultation $T$ (MY27--28) & no & no & no & no & no \\
Occultation $p_s$, held blocks (MY24--28) & no & no & no & no & no \\
Curiosity $p_s$, MY32 holdout & no & no & no & no & no \\
\midrule
\multicolumn{6}{l}{\textit{Validation sets}} \\
MCS $T$, MY32 holdout & no & no & no & yes & yes \\
MCS dust and ice, MY32 holdout & no & no & no & yes & partly \\
MCS dust, held MY28 storm orbits & no & no & no & yes & partly \\
Rover opacity, held site-years & no & no & no & MY32 only & no \\
Curiosity winds, MY32 & no & no & no & yes & no \\
Phoenix winds, MY29 & no & no & no & yes & no \\
\midrule
\multicolumn{6}{l}{\textit{Reference}} \\
MGS occultation $T$, other seasons & no & no & yes & no & no \\
\bottomrule
\end{tabular}
\end{table}

\section{The model}
\label{sec:method}

\begin{figure}[t]
\centering
\figslot{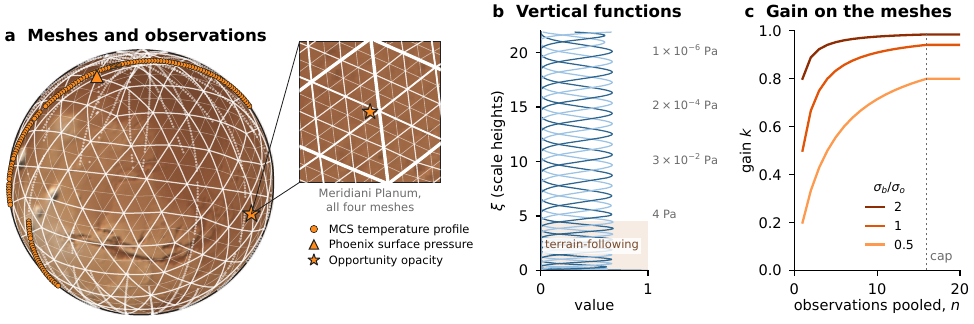}{the pieces of one cycle.}
\caption{The pieces of one MarsNet cycle are drawn from the model itself. (a) The coarsest two of the four nested icosahedral meshes (bold: about 1000~km node spacing; thin: about 500~km) over MOLA topography, with the observations of one real two-hour bin from the training data (MY29, $L_s=98^\circ$, 12:00 Mars coordinated time). Orange dots are the 271 MCS temperature profiles of the bin, the triangle is the Phoenix lander's surface pressure, and the star is Opportunity's column opacity; white dots are the MCS profiles of the rest of that sol. The inset shows all four meshes (finest about 120~km) at Meridiani Planum, Opportunity's landing site. In each cycle the state is advanced two hours, read at these places and times before they are used (where the training loss is taken), and then updated. (b) The 32 cubic B-spline functions that make up every column, in the hybrid coordinate $\xi$ (Section~\ref{sec:state}). The shaded layer below $\xi = 4.5$ blends from terrain-following levels to pressure surfaces. Above it, $\xi = \ln(610\,\mathrm{Pa}/p)$, which gives the pressures listed along the right y-axis. (c) The Kalman-shaped gain on the meshes for three ratios of prior spread to observation error, as more observations are pooled; the cap ($n \le 16$) keeps a dense swath from overwhelming the prior.}
\label{fig:cycle}
\end{figure}

MarsNet is a sequential, learned assimilation model (Figure~\ref{fig:cycle}). Every two hours of Mars time, it advances its state, predicts the observations of that two-hour window from the advanced state (the prior), and then uses the observations to correct the state. Analysis and forecast are the same run, and without observations the cycle is an autoregressive forecast model of the kind used in learned weather forecasting \citep{lam2023graphcast,price2025gencast}. Training adjusts both the forecast and the update to predict each observation from the prior (Section~\ref{sec:trainmain}), so the model learns how to use observations as well as how to predict them.

\subsection{A physical state}
\label{sec:state}

The state holds, in physical units, temperature, surface pressure, surface temperature, the CO$_2$ ice on the surface, winds, and the mixing ratios of dust, water ice, and water vapor. Its finest part lives on four nested icosahedral meshes with node spacings of about 940, 470, 235, and 120~km (Figure~\ref{fig:cycle}a), and above each node every layered quantity is a vertical profile given by 32 cubic B-spline functions (Figure~\ref{fig:cycle}b). Because a column can be evaluated at any pressure, each instrument is read at its own pressures, and no observation is interpolated onto model levels. The state is the sum of five parts that each handle one time scale:
\begin{itemize}[nosep,leftmargin=1.5em]
\item the model's climatology, fixed fields fitted to the observations of the training years, except those held out, as a function of position, height, season, and local time (Appendix~\ref{app:state});
\item a slowly varying bias field \citep{dee2005bias} for the steady offset between the cycled analysis and the model climatology;
\item the large-scale, slowly varying anomalies of temperature and the dust column, such as the departures of a given year from the climatology, carried by LIMs \citep{penland1995lim};
\item a local-time part, of temperature only, which carries the departure of each daily cycle from the climatology, such as a change in the thermal tides, to the same local time of later sols; and
\item finer departures on the meshes, of every quantity, which hold what the other parts leave, down to the smallest scales the meshes resolve.
\end{itemize}
Without data, every anomaly part decays at its own fitted or learned rate, so the state relaxes to the model's climatology, whose error is known, and only that climatology and the bias field may hold the long-term mean (Appendix~\ref{app:state}). MarsNet carries a small ensemble of such states (two members in training and four in evaluation), whose members differ by random perturbations added at every step and together give the model's stated uncertainty with the exact covariance of the LIMs (Appendix~\ref{sec:training}).

\subsection{Advancing the state}
\label{sec:advance}

Zonal means and the largest waves on Mars persist for many sols, while smaller scales lose their memory within a sol \citep{rogberg2010predictability, navarro2017da}, so MarsNet advances each scale separately. Three LIMs fitted to the anomalies of the training years carry the largest scales of temperature from sol to sol, their changes within a sol, and the dust column with their full uncertainty \citep{penland1995lim}; the two temperature LIMs are constrained so that no anomaly grows without observations. A local-time part carries each departure of the daily cycle of temperature to the same local time of later sols, and each quantity on the meshes decays and rotates with a hidden partner at learned rates (Appendix~\ref{app:advance}). Physical tendencies, the rates of change from named physical processes, are computed on the 235~km mesh and move and couple the state: transport in flux form by a balanced wind and a mean overturning, diabatic heating by dust and water ice, a surface energy balance with CO$_2$ ice, the exchange of CO$_2$ with the polar caps, and the lifting, sedimentation, and nucleation of dust and ice (Appendix~\ref{app:advance}).

A graph neural network reads the state of the meshes, the clock, and the surface maps, passes messages between neighboring nodes across all meshes, as in GraphCast \citep{lam2023graphcast}, and returns a context for each column. Each physical term is a small network that reads this context and the local state and sets the coefficients of a fixed physical form, whose sign and limits follow the physics (Appendix~\ref{app:advance}). Beyond these named terms, the network may write a small residual tendency of temperature, surface pressure, dust, ice, and vapor, bounded to at most 1~K per sol for temperature, 0.5\% per sol for surface pressure, and about 10\% per sol for the mixing ratios of dust, ice, and vapor. Each term and residual starts at zero, so the untrained model is a classical filter, which forecasts with fixed, fitted models and corrects toward observations with weights set by their known errors. NeuralGCM takes another route, coupling a full dynamical core, itself a strong physical constraint, with a learned tendency that can take any form \citep{kochkov2024neuralgcm}. MarsNet has no dynamical core, but learning mostly sets the strength of named processes. As a result, each tendency can be examined on its own, its sign is fixed as demanded by physics, and the residual remains bounded where observations are sparse. Some of these restrictions encode physical knowledge, such as the signs of lifting and sedimentation, and others keep the model stable at some cost to what it can represent (Appendix~\ref{sec:stability}).

Winds are not used as training targets, so each ensemble member's wind is diagnosed from its own temperature and surface pressure: in gradient-wind balance away from the equator, in a balance with linear friction near it, and turned by a surface drag in the boundary layer (Appendix~\ref{app:advance}).

\subsection{Using observations}
\label{sec:update}

Each observation is compared with the prior through an observation operator that simulates what the instrument would see, and the difference between the observation and this prediction, the innovation, is what the update corrects. Each observation carries its stated error times a bounded learned factor (Appendix~\ref{app:update}). Before assimilation, we attempt to correct the TES and MCS temperatures for their instrument biases with a fixed table (Section~\ref{sec:eval}), not learned in training, so that the correction does not hide the model errors that the evaluation is meant to show (Appendix~\ref{sec:tbias}).

A Kalman update moves the prior toward each observation by a fraction, the gain, that weighs how uncertain the prior is against how noisy the observation is \citep{Kalman1960}. A confident prior changes little, and an uncertain prior moves most of the way to the observation. The update proceeds in a fixed order, starting with the bias field, which is corrected by an online estimate from all assimilated observations, together with one constant offset for each of the Spirit and Opportunity camera sites. Because the LIMs are linear with a known covariance, their update is the exact Kalman update for the LIM patterns, applied to the slowest patterns first, although it ignores their covariance with the other parts. The meshes are too large and too nonlinear for an exact filter, so the meshes, the local-time part, and a column correction are then updated together from the innovation that the LIMs leave. Each part takes a share $\sigma_i^2$ of the prior variance $\sigma_b^2$ at the observation, as in multiscale analysis schemes \citep{barnes1964,li2015msda}, initially set by how much of the observed variance persists at its scales, and is corrected by
\[
\Delta x_i = g_i\,k_i\,\overline{d}, \qquad k_i = \frac{\sigma_i^2}{\sigma_b^2 + \sigma_o^2/n},
\]
where $\overline{d}$ is the weighted mean of the nearby innovations, $\sigma_o$ the observation error, $n$ the effective number of nearby observations, capped at 16 so that a dense swath of data cannot overwhelm the prior (Figure~\ref{fig:cycle}c), and $g_i$ an opening between 0 and 1 that we add to the Kalman form. The column correction is an optimal interpolation of each column on the finest mesh, which adds the detail of a temperature profile that the meshes cannot resolve, and the local-time part then takes a further share of what all parts leave.

If the later parts also saw the full innovation, they and the LIMs would correct the same large-scale error twice, so the later parts see the innovation that the LIMs leave, less a learned fraction $1 - F$ of its large-scale part, where $F$, learned between 0 and 0.75, reaches 0.60 in the released model. Only the inputs of this Kalman filter are learned: how nearby observations are weighted, how the prior variance is shared among the parts, the fraction $F$, the opening of each mesh, and a bounded factor on each instrument's stated error. Because the shares add up to the prior variance and each opening is at most one, the corrections of all parts after the LIMs together move the prediction at an observation at most the Kalman fraction $\sigma_b^2/(\sigma_b^2 + \sigma_o^2/n)$ of the way toward the weighted mean departure $\overline{d}$ of the observations near it, which keeps them from overshooting the observations, and because $\overline{d}$ is a weighted mean, no correction appears where no observation disagrees with the prior.

\subsection{Training}
\label{sec:trainmain}

Every observation is scored on the prior, before it is assimilated, so every training target is a forecast. The loss is the fair continuous ranked probability score (CRPS) of the ensemble \citep{gneiting2007scoring, ferro2014fair}, which includes the observation error. Training runs the cycle through the training years in chunks of 3.5 sols, each carrying its state forward from the last, and withholds the observations of random spans from the update, so one loss trains both the analysis and forecasts of up to 3 sols. The gradient passes back through the updates and forecasts of each chunk, so an update is rewarded for how well the state it leaves predicts the observations that follow, rather than for how closely it fits those it has just used. The released model is that of step 500 (Appendix~\ref{sec:training}).

Unlike GraphDOP \citep{alexe2024graphdop} and the observation-only reanalysis of \citet{lean2026obsreanalysis}, end-to-end networks trained on a far denser observing system, MarsNet places its network inside a fixed assimilation structure. Learning inside an assimilation scheme has precedents in KalmanNet, which learns the gain of a Kalman filter \citep{revach2022kalmannet}, in auto-differentiable ensemble Kalman filters, which learn the forecast model by training through the filter on observations \citep{chen2022adenkf}, and in 4DVarNet, which learns a variational scheme and its solver \citep{fablet2021varda}, although these have mostly been shown on simulated systems or trained against reference fields. MarsNet learns its forecast and parts of its update through the filter from the real observing record alone.

\section{Results on held-out observations}
\label{sec:results}

We compare MarsNet with MACDA, OpenMARS, and EMARS on the held-out observations that none of the products used (Table~\ref{tab:skill}). Each reanalysis is interpolated to every observation bilinearly in latitude and longitude, linearly in log-pressure, and linearly in time between its outputs, two-hourly for MACDA and OpenMARS and hourly for EMARS. MarsNet is read at the same positions from its prior, a forecast of at most two hours made before the observations of that window are assimilated, and its predicted distribution includes the observation error. For EMARS we use the analysis ensemble mean for temperature and surface pressure, as \citet{greybush2019emars} recommend, and the one-hour background forecast for dust, for which EMARS provides no analysis. We score only the observations that all products cover. For the uncertainties, we use a bootstrap that resamples whole sols, because observations within a sol are correlated.

We score MarsNet on its own on the observations that the reanalyses assimilated, such as the MCS profiles of MY32, by its CRPS skill against two references. The skill is one minus the ratio of MarsNet's CRPS to that of the reference, so it is positive where MarsNet is more accurate. The first reference, the reference climatology, is fitted for each scored set to the observations of the training years, in bins of season, latitude, height, and day or night; it is fitted separately from the model's climatology and more coarsely, with no dependence on longitude and only day and night in local time, so MarsNet is not scored against its own fallback. The second is a damped persistence of the last observed value. The skill is reported by the time since the last assimilated observation, from hours to 31 sols, and by season, latitude, pressure, local time, and dust state. MCS upper limits are scored with the part of the CRPS above each limit. A forecast is calibrated when, over many forecasts, the observations fall within its predicted distribution as often as that distribution implies, so that the spread it predicts matches the error it makes (Figure~\ref{fig:influence}f).

\begin{table}[t]
\centering
\small
\caption{MarsNet and the reanalyses are scored on held-out observations that no product assimilated, by root-mean-square error and, in parentheses, mean bias (product minus observation; Table~\ref{tab:indep}). The lowest RMSE in each row, including ties, is in bold. $N$ is the number of observations. Temperature (K) is by height above the surface in scale heights, surface pressure (Pa), and dust as column opacity at 880~nm. Opportunity's opacity of MY32 was read to pick the model (Table~\ref{tab:indep}), so its row is a validation set rather than an independent test. OpenMARS and EMARS have no output for MY27 $L_s = 180$--$270^\circ$, so the held MGS seasons scored are those of MY25 and MY26. Above about three scale heights, occultation temperatures depend increasingly on the temperature assumed at the top of the retrieval. EMARS is the most accurate reanalysis in the lowest three scale heights and the least accurate above them, so pooled over all heights the ranking changes: against the MEX occultations the errors of MACDA, OpenMARS, and EMARS are 9.2, 8.5, and 9.1~K, and against all MGS occultations of MY24--27 they are 7.9, 7.5, and 8.4~K, where EMARS is also less accurate in MY24 and MY26.}
\label{tab:skill}
\begin{tabular}{llrcccc}
\toprule
Set & Height & $N$ & MarsNet & MACDA & OpenMARS & EMARS \\
\midrule
MGS occultation $T$, held seasons & 0--1 & 7,334 & \textbf{4.2} ($-$1.3) & 5.4 (3.3) & 5.3 (3.4) & 4.4 (2.7) \\
 & 1--2 & 8,198 & \textbf{2.6} ($-$0.9) & 3.2 (0.9) & 2.8 (0.4) & \textbf{2.6} (0.2) \\
 & 2--3 & 7,506 & 4.0 (1.3) & 4.6 (1.9) & 4.3 (1.3) & \textbf{3.9} (0.4) \\
 & 3--4 & 7,028 & 5.7 (1.9) & 5.7 (2.6) & 5.5 (1.9) & \textbf{5.2} (0.9) \\
 & 4--5 & 3,627 & 6.9 (2.6) & \textbf{5.5} (2.2) & 5.9 (1.2) & 5.6 ($-$0.7) \\
MEX occultation $T$ & 0--1 & 2,256 & 7.1 ($-$0.1) & 7.0 (1.8) & 7.2 (1.8) & \textbf{5.7} ($-$0.1) \\
 & 1--2 & 2,444 & 5.0 (1.6) & 5.8 (1.8) & 5.8 (1.2) & \textbf{4.9} (0.5) \\
 & 2--3 & 2,256 & \textbf{6.9} (0.2) & 8.0 (0.4) & 7.0 (0.1) & \textbf{6.9} ($-$1.5) \\
 & 3--4 & 2,134 & 11.3 ($-$1.5) & 12.0 ($-$1.5) & \textbf{11.0} ($-$2.5) & 11.9 ($-$4.0) \\
 & 4--5 & 1,778 & 12.1 ($-$4.2) & 12.2 ($-$4.3) & \textbf{10.8} ($-$3.6) & 14.0 ($-$8.7) \\
Occultation $p_s$, held blocks & surface & 2,827 & \textbf{17.4} (2.0) & 35.4 ($-$22.4) & 22.7 (6.4) & 68.5 (59.1) \\
Curiosity $p_s$, MY32 holdout & surface & 661 & \textbf{9.7} ($-$5.6) & 46.3 ($-$36.9) & 19.9 (13.9) & 33.3 (27.7) \\
Opportunity dust opacity, MY32 & column & 718 & \textbf{0.23} ($-$0.16) & 0.30 ($-$0.27) & 0.33 ($-$0.29) & \textbf{0.23} ($-$0.15) \\
\bottomrule
\end{tabular}
\end{table}

\subsection{Temperature}

Against the held MGS occultations, MarsNet's error near the surface is 4.2~K, compared with 5.4~K for MACDA, 5.3~K for OpenMARS and 4.4~K for EMARS (Table~\ref{tab:skill}). In the lowest three scale heights, MarsNet is more accurate than MACDA and OpenMARS by more than the uncertainty of the difference and level with EMARS. Near the surface, its advantage over EMARS comes from its smaller bias, and with the mean bias removed its error is 4.0~K, against 4.0--4.3~K for MACDA and OpenMARS and 3.5~K for EMARS. Above three scale heights it is less accurate than EMARS, and at 4--5 scale heights less accurate than all three, by 1.0--1.4~K. Its bias changes sign with height, from $-$1.3~K near the surface to +2.6~K at 4--5 scale heights, so the model is too warm where the occultations depend most on their upper boundary and where observations constrain it least (Figure~\ref{fig:constraint}). Above three scale heights, nearly all held occultations are north of 60$^\circ$N, and MarsNet is too warm by 6--8~K in the daytime profiles and unbiased at night on average, although too cold in MY25 and too warm in MY26 by about 3~K; EMARS is also too warm by day, but by less. The reference climatology, fitted to the MCS temperatures of MY28--31 alone, is more accurate at these heights than any of the four products (3.9 and 3.5~K at 3--4 and 4--5 scale heights), so the observing record holds this information and MarsNet loses it, possibly in how it carries the daily cycle of temperature aloft in years when only the TES nadir profiles, which reach about 10~Pa, are assimilated. Against the MEX occultations, which include an offset of about 1~K between the MGS and MEX retrievals near the surface, the errors near the surface are 7.1~K for MarsNet and 5.7--7.2~K for the reanalyses. EMARS is the most accurate there, whereas at 4--5 scale heights MarsNet is more accurate than EMARS. Otherwise the few MEX profiles do not separate the products. The MGS occultations outside the held seasons are the reference of the temperature correction, so they check that reference rather than skill; against them, MarsNet's error near the surface is 5.6~K, with a bias of $-$2.2~K.

On the held MCS profiles of MY32, MarsNet's temperature skill against the reference climatology is 0.29 in the 4-hour windows, where assimilated profiles are hours away, and 0.24 in the 5-sol blocks. At the same time since the last profile, the two holdouts give nearly the same skill, so the skill is set by how long the model has gone without MCS rather than by which holdout is scored. Skill falls slowly with the time since the last assimilated profile, from 0.30 within six hours to 0.23 after 3--7 sols and 0.20 after 7--10 sols, and it stays above that of damped persistence at every age, with a skill against persistence of 0.19--0.23. By height, skill is lowest above 1~Pa (0.18 after 3--10 sols) and highest at 10--100~Pa (0.29). The reanalyses assimilated these profiles, so they are not compared here.

\begin{figure}[t]
\centering
\figslot{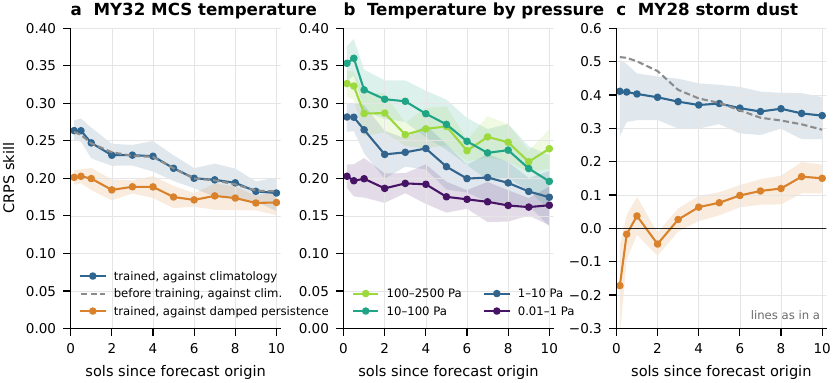}{free forecasts of MY32 and of the MY28 storm, trained and untrained.}
\caption{MarsNet forecasts with every observation after the origin (sol 0) withheld, to show how long its skill lasts without new data. (a) CRPS skill for the held MCS temperatures of MY32, from 218 origins, against the reference climatology for the trained model (blue) and the same model before training (gray), and against damped persistence of the last observed value for the trained model (orange). (b) The trained model against the reference climatology by pressure. (c) As in (a), but for the MCS dust extinction in the held orbits of the MY28 global dust storm, from 28 origins. Between 49 and 130 origins contribute at each time in (a) and (b), and 15--24 in (c). Shading is the 5--95\% interval from a bootstrap over origin sols.}
\label{fig:free}
\end{figure}

To test the forecast on its own, we also ran MarsNet from 218 origins, one every 3 sols of MY32, with every observation after the origin withheld, and scored the held MCS profiles against the same references (Figure~\ref{fig:free}). The skill against the reference climatology falls from 0.26 after four hours to 0.23 after 3 sols and 0.18 after 10 sols, and the forecast stays ahead of damped persistence throughout, with a skill against it of 0.20 falling to 0.17. Skill is highest at 10--100~Pa within the first sol and lowest above 1~Pa at every lead. The predicted spread stays at about 0.9 times the error over the 10 sols, and the forecast drifts warm by 0.4~K. From 28 origins in the MY28 global dust storm, the dust forecast keeps a skill of 0.34--0.41 against the reference climatology, but it overtakes damped persistence only after about 3 sols, and the model holds too little dust in the storm.

\subsection{The influence of an observation}

\begin{figure}[tbp]
\centering
\figslot{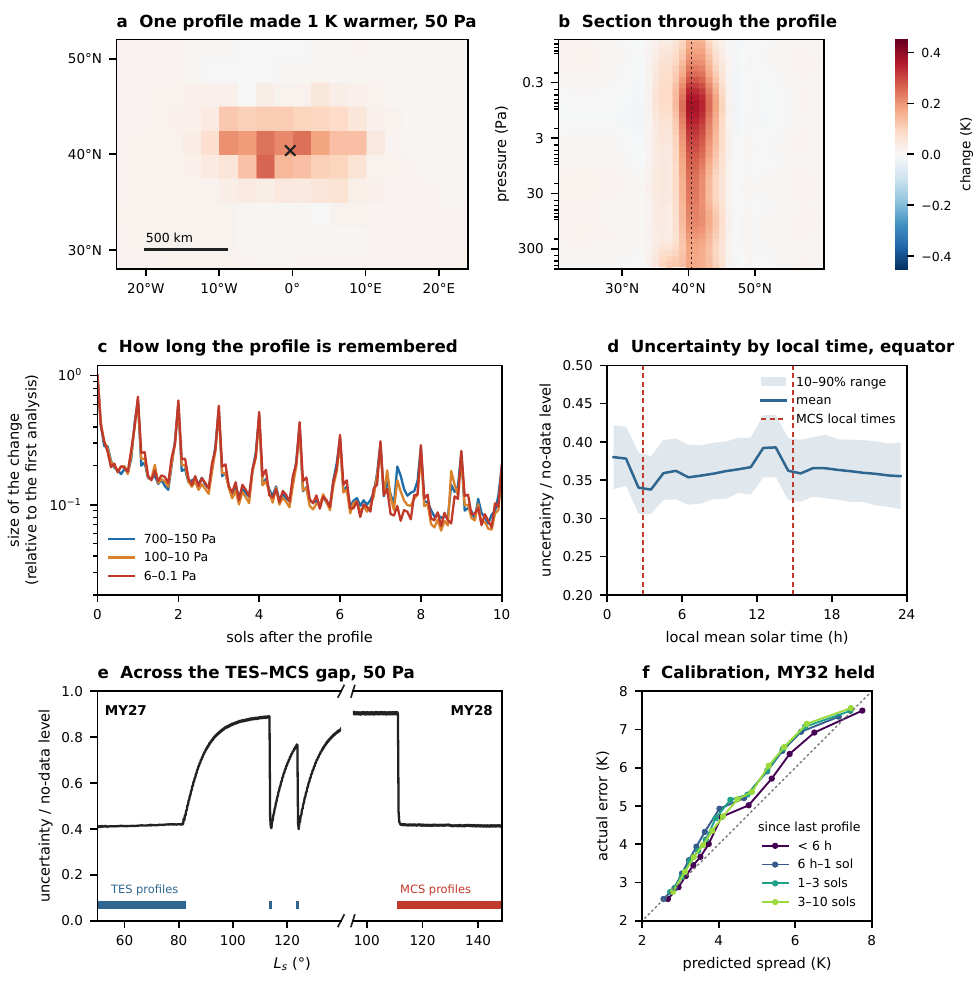}{What an observation does.}
\caption{One MCS temperature profile is made 1~K warmer, to show how a single observation changes MarsNet's state and how long the change lasts. (a) The change of the analysis at 50~Pa (cross: the profile; MY31, $L_s = 175^\circ$, 14:40 local mean solar time), and (b) a latitude-pressure section through it. (c) The root-mean-square size of that change over the following 10 sols, carried with all other observations, in three pressure bands. (d) The model's temperature uncertainty at 50~Pa within 4$^\circ$ of the equator, relative to its level with no observations, by local mean solar time over ten sols; the line is the mean over longitude and time, the shading the 10--90\% range, and the dashed lines mark the local times of the MCS profiles. (e) The same ratio averaged over 60$^\circ$S--60$^\circ$N as TES ends in MY27 (left) and MCS begins in MY28 (right), with about a Mars year without temperature profiles between them; ticks mark the sols with temperature profiles. (f) Actual against predicted error of the held MY32 MCS temperatures, in bins of predicted spread, by time since the last assimilated profile; temperatures more than 3~K below the CO$_2$ frost point are left out.}
\label{fig:influence}
\end{figure}

Because every update is a Kalman update with its own stated uncertainty, the influence of each observation on MarsNet can be followed directly (Figure~\ref{fig:influence}). To demonstrate, we run the assimilation cycle twice from MY31 $L_s = 175^\circ$, once with one MCS temperature profile at 40$^\circ$N made 1~K warmer. The analysis moves about a quarter of the way toward the warmer profile, somewhat less than the Kalman ratio of its prior and observation errors (0.47), and the change falls to $1/e$ within about 200~km in latitude and 400~km in longitude (Figure~\ref{fig:influence}a). It fills the heights the profile sampled (Figure~\ref{fig:influence}b). Carried forward with all other observations, most of the change is gone within a few hours as the next profiles arrive, but part of it returns at the same local time on each later sol, through the local-time part of the state, and that part decays with an e-folding time of about 7 sols (Figure~\ref{fig:influence}c). The model's own temperature uncertainty near the equator stays at about a third of its level with no data. It is lowest, at 0.34 of that level, near 3~h, one of the two local times that MCS observes, and highest, near 0.39, in the hours before the MCS local times, consistent with the local-time part carrying each profile to the same local time of later sols (Figure~\ref{fig:influence}d). After the last regular TES profiles of MY27 the uncertainty roughly doubles, from 0.42 to 0.88 of the no-data level within about 30--40 sols, and when MCS begins in MY28 it falls from 0.90 to 0.43 within a sol (Figure~\ref{fig:influence}e). On the held MY32 profiles, leaving out temperatures more than 3~K below the CO$_2$ frost point, this uncertainty ranks the errors correctly but is slightly too narrow, with a ratio of predicted spread to error of 0.96 within hours of an assimilated profile and 0.91 after one to ten sols, and 88--89\% of the errors fall inside the 90\% interval (Figure~\ref{fig:influence}f).

\subsection{Surface pressure, dust, and winds}

Against the held occultation surface pressures of MY24--28, MarsNet's error is 17~Pa, compared with 23--68~Pa for the reanalyses, and against Curiosity's pressures in MY32 it is 10~Pa, compared with 20--46~Pa, with a mean bias of $-$5.6~Pa. With the mean biases removed, the errors at Curiosity are 7.9~Pa for MarsNet and 14--28~Pa for the reanalyses. MarsNet trains on lander and occultation pressures from other sols and years, whereas the reanalyses do not assimilate surface pressure, so part of this advantage reflects the training data rather than the model alone.

In the held orbits of the MY28 global dust storm, the dust skill against the reference climatology is 0.41, but damped persistence is slightly more skillful (skill against persistence $-$0.06), and the predicted spread there is not calibrated. At Opportunity in MY32, held out for the whole year, the error of the column opacity is 0.23, level with EMARS and below the 0.30--0.33 of MACDA and OpenMARS. All four products are too clear (Table~\ref{tab:skill}), but the reference climatology of the site is more skillful (skill $-$0.18). Between storms, on the held MCS profiles of MY32, the dust extinction does not beat the reference climatology at any age (skill $-$0.04 within six hours of an assimilated profile), and damped persistence beats the model at every age. The losses are largest at the heights where MCS reports only an upper limit, where the model holds too much dust; where dust is detected, the model is level with persistence after 1--10 sols and worse within the first sol. The ice extinction is slightly worse than the reference climatology at every age ($-$0.03 to $-$0.05) and level with persistence, and the model predicts too few exceedances of ice.

Winds are only diagnosed, and the friction near the equator was chosen by comparison with Curiosity's winds of MY32. Against them, the predicted near-surface wind speed is far worse than the reference climatology of the station (CRPS skill $-$1.14), the zonal and meridional winds are worse ($-$0.18 and $-$0.06), and at Phoenix in MY29 all three are worse ($-$0.36 to $-$1.06), so we do not claim skill for the winds.

\subsection{Stability}

Cycled through each of 9 Mars years, each from its own start, the released model stays within its stability bounds, and its stated temperature uncertainty is close to calibrated in the MCS years but not when only TES is assimilated (Appendix~\ref{sec:stability}).

\subsection{What training and the structure contribute}
\label{sec:contrib}

Cycled through MY32 with its weights before training, MarsNet scores nearly as well as the trained model on most held sets, which suggests that much of its skill comes from its structure and fitted tables, themselves built from the observations. Training lowers the error of the held occultation surface pressures from 28 to 17~Pa and slightly narrows the spread of the temperatures, from about 1.0 to about 0.9 times their error, while the skill for the MCS temperatures changes by less than 0.01. It lowers the skill for dust, likely because the trained model gives little weight to observations far from its prior and the model's dust is biased low against MCS, so the MCS aerosol profiles do little to correct it. Without observations, the temperature forecasts of the model before and after training differ in skill by less than 0.01 at every lead (Figure~\ref{fig:free}a).

To test the importance of each part of the update, we removed it from the trained model. We scored the prior temperatures, before each observation is assimilated, in four 10-sol windows of MY28--31 and two of MY25--26. These are training years, so the comparison ranks variants of the model but is not an independent test. Fixing $F$ at 0 or at 1 instead of its learned value raises the mean squared error by 5\% and 1\% in the MCS years and by 42\% and 9\% in the TES years, while fixing it at its starting value of 0.5 changes it by less than 1\%, so the split matters but its change in training does not. Removing the update of the LIMs raises the error by 6\% and 37\%. None of these changes the error of the MCS dust or ice by more than 1\%, although without the update of the LIMs the TES dust columns are predicted 6\% better, and in each case the model stays stable.

\section{Discussion}
\label{sec:discussion}

MarsNet shows that a useful estimate of the Martian atmosphere can be learned from observations alone when the learning is confined within a structure suited to sparse data. MarsNet's forecast and update are trained together to predict each observation before it is used, with no gridded atmospheric state as a target. Near the surface, its temperatures match or improve on those of the reanalyses, its surface pressures are the most accurate of the four products, and its free forecasts of temperature stay ahead of persistence for at least ten sols. For the MCS temperatures, its stated uncertainty is close to calibrated, even in its forecasts, and the effect of each observation can be followed through the state (Figure~\ref{fig:influence}). Much of its temperature skill appears to come from the structure fitted to the observations, while training mainly improves the surface pressures (Section~\ref{sec:contrib}). 

MarsNet is also inexpensive. Training requires about 13 hours on one node, and cycling one Mars year takes about one hour on one GPU. The whole record can be reprocessed within a day whenever a retrieval is revised or a new year of observations arrives. Below we set out its limits and then the uses that this speed and transparency make possible.

The fully independent temperatures, the held MGS seasons and the MEX occultations, cover only MY24--28, and the MEX set is small. Our scored sets contain no occultation temperatures of MY32, so the temperature skill of the validation year is measured against MCS profiles held out of the assimilation, which share the retrieval errors of the profiles that were assimilated around them. The MY32 holdout also served to stop training and choose the released run among a few candidates, so its skill may be slightly optimistic, although the held occultations, on which the products are compared, played no part in that choice.

Because MarsNet has no GCM, it falls back to the model's climatology where observations are missing, as in the gap between TES and MCS, rather than to a GCM's dynamics. The orbiters sample only a few local times, so most of the daily cycle rests on the dynamics, the model's climatology, and a few landers. Winds are constrained only through balance and transport, and we do not claim skill for them. The stated uncertainty is checked only at each observation, so its calibration does not show whether the ensemble represents errors in the position of waves or in transport. In the TES years MY24--27 it is not calibrated (Appendix~\ref{sec:stability}), so the stated uncertainty of the fields in those years should be used with caution.

The model's CO$_2$ frost floor assumes pure CO$_2$, but in the southern polar night non-condensable gases are enriched \citep{sprague2004ar}, and the air can be supersaturated \citep{hu2012co2}. Some of the observed temperatures below the floor may be real, but the model cannot reach them regardless of where in the atmosphere they occur. 1.7\% of the held MCS temperatures of MY32 at pressures greater than 10~Pa, most of them south of 60$^\circ$S, lie more than 3~K below the frost point. In training, the scores of the MCS aerosol profiles and of their upper limits pull against each other, so training improves the upper limits at the cost of the profiles. Temperature and surface pressure, which are observed directly or recur strongly, are where the model is strongest, whereas the aerosols need transport, sources, and sinks to work together. The restrictions that keep the model stable also limit what it can represent. The temperature LIMs cannot let any anomaly grow, even briefly, and the large scales drive the finer departures but not the reverse.

The speed of training and of the assimilation cycle makes MarsNet a tool for experiments with the observing record itself. Adding a data set, revising a retrieval, or withholding an instrument needs only a new run of the cycle, or for a new kind of observation a retraining, and its effect can be scored on the same held-out observations within a day. Observing system experiments of this kind, which on Earth guide the design of the observing network, can then be run routinely on the real Martian record, by withholding one instrument, one range of local times, or one season and measuring how much the estimate degrades. Because each correction is a Kalman update with its own stated uncertainty, MarsNet also shows how strongly the record constrains its estimate (Figure~\ref{fig:influence}). These measures describe how MarsNet uses the observations, and a data-denial experiment can test them by checking that withholding an instrument degrades the held-out scores as much as the stated uncertainty predicts. Tested in this way, they could help to plan future missions, for example by estimating how much an orbiter sampling other local times or a network of surface stations would add.

MarsNet also offers a reference for intercomparison that rests on different assumptions. It assimilates the same retrievals as the reanalyses, but because no GCM or reanalysis enters it, its disagreements with the reanalyses and with free-running GCMs suggest where a model's physics rather than the observations sets the answer, although differences in retrieval biases, sampling, and assimilation can also contribute. MarsNet's held-out sets let every product be scored on equal terms. Its temperature correction gives a consistent account of how TES, MCS, and the occultations differ by height, local time, season, and year, which could support future revisions of their retrievals. The model can also be compared with GCMs process by process. Because its learned terms set the strength of named processes rather than an arbitrary tendency, the learned rates of dust lifting and sedimentation, the heating by dust and water ice, and the decay of the large-scale patterns can be set beside the corresponding parameterizations, keeping in mind that a learned rate can absorb errors elsewhere in the model. Where the bounded residual tendency stays large, it marks a candidate for a process that the named terms do not yet describe.

We see the largest opportunity in incorporating richer dynamics. Where observations are missing, MarsNet now falls back to its climatology, whereas a dynamical core inside the same structure, corrected only by observations, could carry the circulation through gaps and constrain the winds and the daily cycle. Each such addition can be tested against the same held-out observations, so hybrid models of this kind, built on physical equations but trained on observations alone, can be developed step by step with every gain measured. A version trained on all years would then provide fields with uncertainties for the whole record, as a common basis for studies of the Martian climate. Even in its present form, MarsNet may serve as a tool for discovery, wherever an estimate of the Martian atmosphere that rests on the observations alone can test what the models assume.

\section*{Code and data}
All observations and surface maps come from their primary sources, mostly the NASA Planetary Data System and the ESA Planetary Science Archive. The MarsNet source code, the weights and configuration of the released model, its fitted tables, its fields, and the held-out observation sets used for scoring will be released openly in the coming months, with a DOI. The code includes the pipeline that downloads every input from these sources and checks it against a recorded checksum, so the training data can be rebuilt from the primary sources, and the scoring that recomputes every number in this paper.

\section*{Acknowledgments}
We acknowledge the creators of the MACDA, OpenMARS, and EMARS reanalyses, which made the comparisons in this paper possible. This work was not supported by a grant.

\section*{Use of AI}
AI agents were used in the training and implementation of MarsNet and in the preparation of this paper. The authors directed the research, reviewed the results, and take responsibility for the content.

{\small
\setlength{\bibsep}{0.25em}
\bibliographystyle{plainnat}
\bibliography{refs}}

\clearpage
\appendix

\section{Model details}
\label{app:model}

\subsection{The state}
\label{app:state}

Each vertical profile is stored as the coefficients of 32 cubic B-spline functions \citep{untch2004fe} (Figure~\ref{fig:cycle}b). Each function is a smooth bump spanning a few neighboring layers, packed closer together near the ground, so the profile is smooth and a correction at one height leaves the rest unchanged. The vertical coordinate $\xi$ is a hybrid of that used in GCMs \citep{simmons1981hybrid}, defined by $\ln p = \ln p_0 - \xi + B(\xi)\ln(p_s/p_0)$ with $p_0 = 610$~Pa. The weight $B$ falls smoothly from one at the surface to zero at $\xi = 4.5$ (about 7~Pa), so the lowest levels follow the terrain and those above $\xi = 4.5$ are surfaces of constant pressure. The scores of Section~\ref{sec:results} are instead binned by scale heights above the local surface, $\ln(p_s/p)$, which equals $\xi$ only where $p_s = 610$~Pa. The physics is computed up to $\xi = 12$ (about 0.004~Pa). An MCS limb profile is read at the heights it sampled, a TES nadir sounding as a weighted average over a deep layer, and a lander barometer at the surface, so every observation constrains the same atmosphere. Water vapor is stored as the logarithm of its ratio to the water vapor of the model's climatology, fitted to the CRISM columns of the training years and limited by saturation over ice.

For temperature, the model's climatology is a smooth fit in harmonics of season, local time, and longitude on a grid of latitude and height, plus a correction fitted to the MCS profiles in bins of latitude, longitude, season, local time, and height, in which bins with few profiles fall back to their zonal mean. The bias field is a smooth function of latitude, height, and season, with the first two harmonics of the daily cycle of temperature, stationary waves of zonal wavenumbers 1--3, and the non-migrating tides of the same wavenumbers. The LIM patterns are fields on their own 10$^\circ$ grid, and the local-time part is a set of daily harmonics on the 235~km mesh (Section~\ref{sec:advance}). All five parts are evaluated and summed at each observation and at each node where the physics is computed. During a forecast the parts interact in only one direction. The LIMs step forward on their own, while the physics reads the sum of all parts and writes its tendencies only into the mesh departures, so the large-scale anomalies shape the finer departures but not the reverse. In the update, every part is corrected from the innovation of each observation against the full state at that observation. The relaxation to the climatology in the absence of data is not prescribed for each gap, because every anomaly component decays exponentially at its own rate, so the state relaxes with the time since the last observation. The anomaly components are kept centered on zero, so the model cannot slowly drift away from the climatology.

\subsection{Advancing the state}
\label{app:advance}

On Earth, linear, damped models are widely used to capture and predict the large-scale anomalies that persist for weeks \citep[e.g.,][]{winkler2001lim}, and the Mars Climate Database likewise represents large-scale variability with a set of patterns, empirical orthogonal functions of its GCM runs \citep{millour2026}. LIMs have served as the forecast model of a Kalman filter for sparse paleoclimate data \citep{perkins2017lim}, and a LIM with a learned correction has beaten both a LIM and a fully learned model when only a century of observations is available \citep{schloer2025hybrid}. On Mars, the largest scales, which persist for many sols, are carried by three LIMs: one for temperature from sol to sol with 200 patterns, one for the sub-daily temperature anomalies in steps of six hours with 200 patterns and one for the dust column with 40 patterns. Each describes anomalies as a set of patterns on a 10$^\circ$ grid, which resolve the zonal means and the largest scales in longitude, and their amplitudes $x$ step forward as $x_{t+1} = A x_t + \epsilon$, with the matrix $A$ and the noise $\epsilon$ fitted to observed anomalies of the training years, those of MCS in MY28--31 for the two temperature LIMs and those of MY24--31 for dust. The fit of the sol-to-sol LIM is pulled toward an equal decay of all patterns, by an amount chosen by fitting on all but one training year and testing on the year left out, and it is used only if it beats persistence at leads of 1--3 sols on held-out MCS profiles of MY32. As fitted, the two temperature LIMs could still let anomalies grow for weeks without observations, as in the gap between TES and MCS, so their matrices are then limited so that, with each pattern scaled by its spread in the training years, no combination of patterns grows from one step to the next. The noise of the sol-to-sol LIM is set to three quarters of the amount that would keep the anomalies at their observed spread, so in a long gap they settle just within it. Half of the sol-to-sol temperature patterns lose their amplitude to $1/e$ within about 15 sols.

The daily cycle of temperature is observed at only a few local times by each orbiter, so it is carried by a separate part of the state. It holds the first four harmonics of the daily cycle at each node of the 235~km mesh and decays with an e-folding time of up to 10.5 sols, so a departure seen at one local time is carried to the same local time of later sols. A memory of the departure at each hour of the day, with e-folding times of 2--25 sols that depend on height, adds detail that four harmonics cannot hold. When the model is run ahead of the last observations, as in a forecast, this memory fades three times more slowly above about 0.6~Pa ($\xi = 7$), which keeps the upper atmosphere from drifting cold within a few sols. The summed amplitude of the harmonics is bounded at 40~K, just above the largest daily departure in the training years, because near the poles, where MCS samples only two narrow bands of local time, the harmonics could otherwise grow without limit.

The departures on the four meshes are advanced more simply. A full linear model coupling every node and quantity would have far more coefficients than the observations can constrain, so in this step each quantity at each node evolves on its own, and the coupling between nodes comes from the physical tendencies described below. Each quantity is paired with a hidden partner, and together the two behave as a damped oscillation that decays exponentially while rotating at a fixed frequency, so a departure can return later in the sol rather than only fade. On each mesh, each pair has one decay rate and one frequency, shared by all nodes and heights, learned in training, and allowed to vary with season, with an e-folding time of at least a quarter sol. The rates for temperature start from those of the sol-to-sol LIM, and for surface pressure, surface temperature, and winds the frequency starts at one cycle per sol, so a departure in the daily cycle recurs at the same local time on the next sol.

Movement and coupling between nodes and between quantities come from physical tendencies, computed on the 235~km mesh. Dust, water ice, water vapor, and heat are carried in flux form, which conserves their mass, by a reference flow made of the balanced wind (below), the mass flux of the surface-pressure tide observed at Curiosity, and a mean meridional overturning. The overturning follows from mass conservation, the zonal-mean heat and momentum equations, and gradient-wind balance of the zonal mean, which together give the mean meridional circulation that keeps the zonal-mean wind and temperature in that balance, driven by the diagnosed solar and infrared heating of dust, ice, and CO$_2$, by the heating of each ensemble member, and by the eddy heat and momentum fluxes of the balanced wind. Temperature also changes through diabatic heating from lofted dust and water-ice clouds and through heat exchange with the surface within the boundary layer. Surface temperature follows a surface energy balance of sunlight, downwelling infrared, and sensible heat over a 22-layer subsurface, with a learned factor on the strength of its tendency. Where the surface holds CO$_2$ ice, its temperature is held at the frost point until the ice has sublimated, and the ice grows and shrinks with the surface energy deficit. Surface pressure follows mass continuity and the seasonal exchange of CO$_2$ with the polar caps. Where the air cools below the CO$_2$ frost point for a given pressure (e.g., $\sim$147~K at 610~Pa), it is held at the frost point, and the condensed mass is removed from the surface pressure. This floor applies to every ensemble member at every step, and the analysis is held no more than 3~K below it. Dust is lifted from the surface, sediments, and nucleates on water ice, and a deficit in the column refills over about two sols. Water ice is transported, sediments, and sublimates. Water vapor changes only by the water that ice takes up as it forms and releases as it sublimates, limited by the mass available. Vapor above saturation is not condensed, and no other process reads the vapor. The divergent part of the wind is damped. Within their fixed forms, dust lifting cannot be negative, nucleation and sublimation switch on below and above the water frost point, boundary-layer heating relaxes the air toward the surface temperature, the heating by dust and ice reads the temperature of the ensemble's reference state and can only damp each member's departure from it, and sedimentation follows Stokes' law with a slip correction, whose rate is learned within a factor of $e$. The model does not include atmospheric CO$_2$ ice or radiative transfer in the atmosphere beyond these heating terms.

The model splits the wind into two parts: a balanced wind computed from the temperature and surface pressure it does observe, and a departure, the difference between the full wind and the balanced wind, which is damped and moves no mass, heat, or tracer. The balanced wind is computed from the full state on a surface pressure that is consistent with each column's own temperature. It is integrated hydrostatically through the column's temperature from a datum pressure fitted to the landers, rather than extrapolated with a fixed scale height, which is meant to avoid spurious circulations over the deepest basins such as Hellas. Its zonal mean is in gradient-wind balance, in which the pressure gradient is balanced by the Coriolis force and the curvature of the flow, and its eddies are in balance at each point \citep{Gent1983, Randel1987, Boville1987, Lieberman1999}. Within 10$^\circ$ of the equator, where the Coriolis force vanishes, the wind is a steady balance with linear friction, blended into the balanced wind by 20$^\circ$. Near the surface, a bulk drag produces an Ekman layer, in which surface friction slows the wind and turns it toward low pressure. Its depth follows the daily cycle of mixing, from about 200~m at night to up to 10~km by day. Each ensemble member has its own balanced wind, which gives the winds their spread.

\subsection{Using observations}
\label{app:update}

For the orbiters, the observation operators read the model's temperature at the height of each retrieved profile point, compute extinction from mixing ratio and air density, and scale columns to 610~Pa. Where MCS reports only an upper limit of extinction, the operator gives the probability that the model's extinction lies below it, and the update pulls the prior down only where it predicts extinction above the limit. For the rovers, the operators convert the model's dust column at the site's pressure to the visible opacity the rovers measure, including the water ice that the rovers also see. For surface pressures from landers and radio occultations, they move the model's surface pressure hydrostatically from the height of its terrain to the height of the observation. The learned factor on each observation's stated error has fixed bounds \citep[cf.][]{desroziers2005diag}, and for MCS temperatures the factor is fixed at 0.55, and the spread of the predicted observation is scaled in each height band by a table fitted to the innovations of the untrained model. The loss and the update use this same error.

The large-scale part removed through $F$ is the part of the innovation that the LIM patterns can represent. With $F = 1$ the untrained model is unstable, and over 100 sols its analysis drifts tens of kelvin from the observations, whereas with $F = 0$ the later parts never see the large scales. In training, $F$ starts at 0.5. Updates are localized in the vertical \citep{gaspari1999localization}, and each observation only changes its own variable, so a temperature observation never moves the dust. The one exception is the semidiurnal pressure tide at Curiosity, which corrects the global dust column through a fitted linear relation, given the strong relationship between rover tides and dust opacity \citep{guzewich2016tides, zuritazurita2022storms, leino2023tides}. A zonal-mean dust correction from the rover opacities shares the prior variance of the parts after the LIMs and reads the whole innovation that the LIMs leave, not the one reduced through $F$.

\section{Instrument temperature biases}
\label{sec:tbias}

\begin{figure}[t]
\centering
\figslot{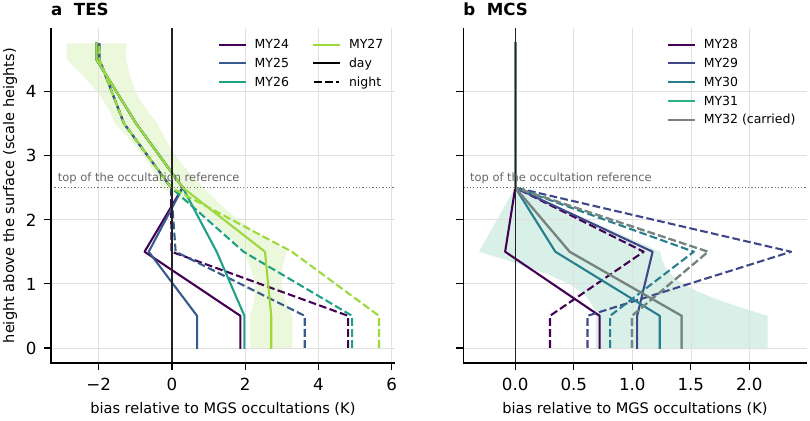}{the TES and MCS temperature corrections against height, day and night, with their 1-sigma bands; one line per Mars year.}
\caption{The instrument temperature correction that MarsNet subtracts before assimilation is shown for (a) TES and (b) MCS against height above the surface in scale heights, by day and by night, at the equator and averaged over the season. Lines are the training years, and shading is the 1$\sigma$ uncertainty of the fit by day for MY27 (TES) and MY31 (MCS); MY32 (gray) has no fit of its own and carries that of MY31. The dotted line marks 2.5 scale heights, above which the occultations no longer constrain the correction.}
\label{fig:tbias}
\end{figure}

TES and MCS temperatures differ from each other and from the radio occultations by a few kelvin, depending on height, local time, and season, and if left in, the differences would give the model a different atmosphere in the TES and MCS eras. The correction is a fixed table (Section~\ref{sec:update}), fitted to collocations of the training years: TES and MGS occultations within the same sols, and TES, MCS, and MGS occultations of different years at the same season, latitude, and local time, whose difference also contains the change between years. The MGS occultations outside the held seasons are the reference (Section~\ref{sec:eval}). For each instrument, the bias is fitted at five heights from 0.5 to 4.5 scale heights and interpolated between them, with a term for each Mars year, a day-night difference, a seasonal cycle, and a dependence on latitude, and the term for each year may differ from the next by a random step, so a sparse year shrinks toward its neighbors and a real drift is kept. Daytime collocations of the dusty season ($L_s = 180$--$360^\circ$) and all collocations of the global dust storms of MY25 and MY28 are left out of the fit. Near the surface, TES is too warm by 0.7--2.7~K by day and 3.6--5.7~K by night, and MCS by 0.3--1.4~K, with an uncertainty of 0.3--0.8~K. Above 2.5 scale heights, where too few occultations reach to constrain the fit, MCS is taken as its own reference and its correction is zero, and the TES correction, tied to MCS through collocations of different years, reaches $-$2~K at 4.5 scale heights (Figure~\ref{fig:tbias}). MY32 is kept out of the fit so that its scores stay independent, and its correction is that of MY31 carried forward with the added uncertainty of one random step. A fixed table cannot adjust to biases that the collocations miss, and with more independent profiles to compare against, such as later radio and solar occultations, future versions may learn part of the correction in training while those profiles keep it from absorbing errors of the model itself.

\section{Training}
\label{sec:training}

For an MCS upper limit, only the part of the predicted distribution above the limit is scored. The score of each kind of observation is divided by the score of the model's climatology, with its stated error, on that kind in the training years, so that each kind counts on the same scale, and is weighted by variable (one for temperature and dust, one half for ice, surface temperature, and surface pressure, and one quarter for water vapor). Rover opacities are assimilated but carry no weight in the loss. The ensemble has two members in training and four in evaluation. The scores of Section~\ref{sec:results} use the CRPS of the predicted distribution of the four members, a mixture of Gaussians that each include the observation error, so they are not the training loss itself. The members differ by Gaussian noise that is added at every step on each mesh, correlated in space over about 1.5 node spacings and in time at the decay rate of each quantity, with a learned amplitude.

We train on chunks of 3.5 sols, carried forward along streams, continuous runs of the cycle of up to 120 sols in the training years. Each step trains on one chunk of each of 16 streams, whose start dates are spread evenly over the seasons. In half of the steps, one span of 3 sols is withheld from the update for every instrument and scored as a free forecast. Observations are also withheld from the update in random windows, of four hours to eight sols for MCS and of half a sol to eight sols for all instruments, so that the model can learn to carry the state through data gaps; unlike the holdout of MY32, these windows serve only as training targets and are not used as tests. We use the AdamW optimizer \citep{loshchilov2019adamw} with a learning rate of $1.4 \times 10^{-4}$, warmed up over 100 steps and decaying slowly thereafter, and train for 500 steps. Training is in double precision, because single precision gave undefined losses. It ran on one node with eight NVIDIA RTX PRO 6000 GPUs for about 13 hours.

\section{Stability}
\label{sec:stability}

Because a reanalysis must stay close to the observations for years without drifting, we cycled the released model through each of 9 Mars years, MY24--32, starting each year from its own initial state, and checked the size of each part of the state against its bound and the statistics of its innovations. In every year the state stays within its bounds; the closest approach is the sol-to-sol LIM in MY28, at 84\% of its bound. In the MCS years MY28--32, the ratio of temperature innovation variance to its prediction is 1.0--1.6, close to one, but in the TES years MY24--27 it ranges from 0.28 to 12.8, so the stated uncertainty is too small in MY24 and MY27 and too large in MY26 and is not calibrated when only TES is assimilated. Three restrictions keep the cycle stable (Appendix~\ref{app:advance}). Without the restriction that the learned heating by dust and ice can only damp, training made that heating amplify temperature anomalies faster than they decay, so a cold anomaly over the pole fed on itself. Without the limit on their growth, the temperature LIMs let anomalies grow in data gaps until their variance reached up to 300 times that of the training years. Without the bound on its amplitude, the daily pattern near the poles, traded between the local-time part and the meshes, reached 76~K, whereas with the bound it peaks at 19~K in the cycled years.

\end{document}